\documentclass[aps,pre,amsmath,amssymb,floatfix,preprint]{revtex4-1}
\usepackage{graphicx}
\usepackage{color,multirow}

\begin{document}

\title{Continuum limit of the vibrational properties of amorphous solids}

\author{Hideyuki Mizuno}
\email{hideyuki.mizuno@phys.c.u-tokyo.ac.jp} 
\affiliation{Graduate School of Arts and Sciences, The University of Tokyo, Tokyo 153-8902, Japan}

\author{Hayato Shiba}
\affiliation{Institute for Materials Research, Tohoku University, Sendai 980-8577, Japan}

\author{Atsushi Ikeda}
\email{atsushi.ikeda@phys.c.u-tokyo.ac.jp} 
\affiliation{Graduate School of Arts and Sciences, The University of Tokyo, Tokyo 153-8902, Japan}

\date{\today}

\begin{abstract}
The low-frequency vibrational and low-temperature thermal properties of amorphous solids are markedly different from those of crystalline solids. 
This situation is counter-intuitive because any solid material is expected to behave as a homogeneous elastic body in the continuum limit, in which vibrational modes are phonons following the Debye law. 
A number of phenomenological explanations have been proposed, which assume elastic heterogeneities, soft localized vibrations, and so on. 
Recently, the microscopic mean-field theories have been developed to predict the universal non-Debye scaling law. 
Considering these theoretical arguments, it is absolutely necessary to
directly observe the nature of the low-frequency vibrations of
amorphous solids and determine the laws that such vibrations obey. 
Here, we perform an extremely large-scale vibrational mode analysis of a model amorphous solid. 
We find that the scaling law predicted by the mean-field theory is violated at low frequency, and in the continuum limit, the vibrational modes converge to a mixture of phonon modes following the Debye law and soft localized modes following another universal non-Debye scaling law. 
\end{abstract}

\maketitle

\section{Introduction}
The low-frequency vibrational and low-temperature thermal properties of amorphous solids have been a long-standing mystery in condensed matter physics.
Crystals follow universal laws, which are explained in terms of phonons~\cite{kettel,Leibfried}. 
Debye theory and phonon-gas theory predict that the vibrational
density of states~(vDOS) follows $g(\omega) \propto \omega^{2}$, the
heat capacity follows $C \propto T^3$, and the thermal conductivity
follows $\kappa \propto T^3$ in three-dimensional systems, which
indeed agree with experimental results ($\omega$ is frequency, and $T$ is temperature). 
Similarly, amorphous solids are characterized by universal laws;
however, these laws are anomalous with respect to those of crystalline solids~\cite{lowtem}.
At $T \sim 10$~K, the heat capacity of amorphous solids becomes larger
than the value for crystalline solids~\cite{Zeller_1971}, which directly reflects the excess vibrational modes around $\omega_\text{BP} \sim 1$~THz~\cite{buchenau_1984}, often referred to as the boson peak (BP).
At $T \lesssim 1$~K, the thermal conductivity increases as $\kappa
\propto T^2$ rather than $\kappa \propto T^3$~\cite{Zeller_1971},
indicating that the vibrational modes are not phonons even at very low frequency $\omega \sim 0.1$~THz~(one order of magnitude lower than $\omega_\text{BP}$).
These behaviours are highly counter-intuitive because any solid material, not only crystalline but also amorphous, is expected to behave as a homogeneous elastic medium in the continuum limit, and its vibrational modes are expected to converge to phonons~\cite{tanguy_2002,Monaco2_2009}.

A number of theoretical explanations for these anomalies have been
proposed, and these explanations substantially differ. 
One
approach~\cite{Marruzzo_2013,Marruzzo2_2013,Mizuno2_2013,Mizuno_2014,Gelin2016}
assumes that an inhomogeneity in the mechanical response at the nanoscale~\cite{yoshimoto_2004,Tsamados_2009,Mizuno_2013} plays the central role.
In this approach, the heterogeneous elasticity equation is solved using the effective medium technique to predict the boson peak and anomalous acoustic excitations~\cite{Marruzzo_2013,Marruzzo2_2013}.

Another approach is the so-called soft potential model~\cite{Karpov_1983,Buchenau_1991,Buchenau_1992,Gurevich_2003}, which is an extension of the famous tunnelling two-level systems model~\cite{Anderson_1972}. 
This theory assumes soft localized vibrations to explain the anomalous thermal conductivity and the emergence of the boson peak. 
Soft localized vibrations have been numerically observed in a wide variety of model amorphous solids~\cite{mazzacurati_1996,Taraskin_1999,Schober_2004,Xu_2010,Lerner_2016} and in the Heisenberg spin-glass~\cite{Baity-Jesi2015}. 
Interestingly, these localized modes are also argued to affect the dynamics of supercooled liquids~\cite{widmer_2008} and the yielding of glasses~\cite{Maloney2_2006}.

Recently, a quite different scenario has been emerging. 
This scenario is based on studies of the simplest model of amorphous solids~\cite{OHern_2003}, which is randomly jammed particles at zero temperature interacting through the pairwise potential,
\begin{equation}~\label{pot-simple}
\phi(r) = \frac{\epsilon}{2} \left( 1 - \frac{r}{\sigma} \right)^2 H(\sigma -r),
\end{equation}
where $H(r)$ is the Heaviside step function and $\sigma$ is the
diameter of the particles. 
When the packing pressure $p$ is lowered, the particles lose their contacts at zero pressure $p=0$, which is called the (un)jamming transition~\cite{OHern_2003}. 

Importantly, the mean-field theory analysis of this model is now
advancing considerably, providing a new way to understand the anomalies of amorphous solids~\cite{Wyart_2005,Wyart_2006,Parisi2010,Wyart_2010,Berthier2011a,Wyart2012,DeGiuli_2014,Charbonneau2014b,Charbonneau2014,Franz_2015,Biroli_2016}.
Previous theoretical~\cite{Wyart_2005,Wyart_2006} and numerical~\cite{Silbert_2005,Silbert_2009} works have clearly established that the vDOS of this model exhibits a characteristic plateau at $\omega > \omega_\ast$, where $\omega_\ast$ is the onset frequency of this plateau.
The relevant region for the low-frequency anomalies of amorphous solids is below this plateau, $\omega < \omega_\ast$. 
First, the effective medium theory assuming marginal stability, i.e., that the system is close to the elastic instability, predicts the characteristic behaviour of the vDOS $g(\omega) = c \omega^2$ at $\omega \ll \omega_\ast$~\cite{Wyart_2010,DeGiuli_2014}.
This prediction differs from the prediction of Debye theory $g(\omega)=A_0 \omega^2$ because the prefactor $c \propto {\omega_\ast}^{-2}$ is considerably larger than the Debye level $A_0 \propto {\omega_\ast}^{-3/2}$. 
Thus, this prediction provides a new explanation of the boson peak in terms of marginal stability~\cite{Wyart_2010,DeGiuli_2014}.
Second, the model is analysed using the replica theory~\cite{Parisi2010,Berthier2011a,Charbonneau2014b,Charbonneau2014,Franz_2015,Biroli_2016}. 
This theory becomes exact in infinite
dimensions~\cite{Charbonneau2014b}; thus, it can be a firm starting
point for considering the problem.
The theory predicts that the transition from the normal glass to the marginally stable glass occurs at a finite pressure $p = p_G$, which is called the Gardner transition~\cite{Biroli_2016}. 
Near the Gardner transition, the theory predicts non-Debye scaling of the vDOS $g(\omega) = c \omega^2$ at $\omega \ll \omega_\ast$~\cite{Franz_2015}, which perfectly coincides with the prediction from the effective medium theory.
Remarkably, in the marginally stable glass phase, the region of this scaling law extends down to $\omega \to 0$, which means no Debye regime even in the continuum limit (the limit of $\omega \to 0$)~\cite{Franz_2015}.
This non-Debye scaling law was recently shown to work at least near $\omega_\ast$~\cite{Charbonneau_2016}. 

Considering these different theoretical arguments, it is absolutely
necessary to numerically observe the nature of the low-frequency
vibrations of amorphous solids and determine the laws that such vibrations obey. 
This task is not as easy as it sounds because the lower the frequency
that we require, the larger the system that we need to simulate. 
Here, we perform a vibrational mode analysis of the model amorphous solid defined by Eq.~(\ref{pot-simple}), composed of up to millions of particles ($N \sim 10^6$), which enables us to access extremely low-frequency modes even far below $\omega_\text{BP}$. 
Then, we investigate the nature of the modes in detail by calculating several different parameters. 
Notably, we find that the non-Debye scaling $g(\omega) = c \omega^2$ is violated at low frequency, and in the continuum limit, the vibrational modes converge to a mixture of phonon modes following the Debye law and soft localized modes following another universal non-Debye scaling law, the $\omega^4$ scaling law, which was recently first observed in Refs.~\cite{Lerner_2016,Baity-Jesi2015} by suppressing the effects of phonons.

\section{Methods}
%
\subsection{System description}
We study three-dimensional (3D, $d=3$) and two-dimensional (2D, $d=2$)
model amorphous solids, which are composed of randomly jammed particles~\cite{OHern_2003}.
Particles $i,j$ interact via a finite-range, purely repulsive,
harmonic potential $\phi(r_{ij})$~[Eq.~(\ref{pot-simple})], where
$r_{ij}=|\mathbf{r}_i - \mathbf{r}_j|$ is the distance between the two particles.
The 3D system is mono-disperse with a diameter of $\sigma$, whereas
the 2D system is a $50\%$-$50\%$ binary mixture with a size ratio of
$1.4$ (the diameter of the smaller species is denoted by $\sigma$).
The particle mass is $m$.
Length, mass, and time are measured in units of $\sigma$, $m$, and $\tau=(m \sigma^2/\epsilon)^{1/2}$, respectively.
To access low-frequency vibrational modes, we consider several different system sizes $N$ (number of particles), ranging from relatively small $N=16000$ to extremely large $N=2048000$.
We always remove the rattler particles that have less than $d$ contacting particles.

Mechanically stable amorphous packings are generated for a range of packing pressures from $p \sim 10^{-1}$ to $10^{-4}$.
We first place $N$ particles at random in a cubic (3D) or a square (2D) box with periodic boundary conditions in all directions.
The system is then quenched to a minimum energy state.
Finally, the packing fraction $\varphi$ is adjusted by compressive deformation (CO) until a target pressure $p$ is reached.
We also study the shear-stabilized (SS) system~\cite{Dagois-Bohy_2012} by minimizing the energy with respect to the shear degrees of freedom.
No differences between the CO and the SS systems have been confirmed for our systems of $N \ge 16000$.
In this paper we present the results obtained from the CO system.

In the packings obtained using the above protocol, the inter-particle forces are always positive $-\phi'(r) > 0$.
For this reason, we refer to this original state as the stressed system. 
In addition to the stressed system, we also studied the unstressed system, where we retain the stiffness $\phi''(r)$ but drop the force $-\phi'(r) = 0$ in the analysis.
Since the positive forces make the system mechanically unstable, dropping the forces makes the original stressed system more stable~\cite{Wyart_2006,DeGiuli_2014,Lerner_2014}.

\subsection{Vibrational mode analysis}
We have performed the standard vibrational mode
analysis~\cite{kettel,Leibfried}; we have solved the eigen-value
problem of the dynamical matrix ($dN \times dN$ matrix) to obtain the eigen value $\lambda^k$ and the eigen vector $\mathbf{e}^k=\left[ \mathbf{e}^{k}_1,\mathbf{e}^{k}_2,...,\mathbf{e}^{k}_N \right]$ for the modes $k=1,2,...,dN-d$ ($d$ zero-$\omega$, translational modes are removed).
Here the eigen vectors are orthonormalized as $\mathbf{e}^k \cdot
\mathbf{e}^l = \sum_{i=1}^N \mathbf{e}_i^k \cdot \mathbf{e}_i^l =
\delta_{k,l}$, where $\delta_{k,l}$ is the Kronecker delta function.

From the dataset of eigen frequencies, $\omega^k = \sqrt{\lambda^k}$ ($k=1,2,...,dN-d$), we calculate the vDOS as
\begin{equation} \label{equofvdos}
g(\omega) = \frac{1}{dN-d} \sum_{k=1}^{dN-d} \delta \left( \omega-\omega^k \right),
\end{equation}
where $\delta(x)$ is the Dirac delta function.

In this work we analysed several different system sizes of $N=16000$ to $2048000$.
We first calculated all the vibrational modes in the system of $N=16000$.
We then calculated only the low-frequency modes in the larger systems of $N > 16000$.
Finally the modes obtained from different system sizes were put together as a function of the frequency $\omega^k$.
We found that the results from different system sizes smoothly connect with each other, which provide the modes information in a wide frequency regime.
The vDOS was calculated from these dataset, and also the results of these different system sizes are presented altogether in the figures.

The present system exhibits a characteristic plateau in $g(\omega)$ at $\omega > \omega_\ast$; $\omega_\ast$ is defined as the onset frequency of this plateau (see the Supplemental Information, Fig.~S1)~\cite{Silbert_2005,Silbert_2009}.
Practically, we determined the value of $\omega_\ast$ as the boson peak position of the unstressed system~\cite{Wyart_2006,Mizuno3_2016}.

\subsection{Phonon and Debye vDOS}
In an isotropic elastic medium, phonons are described as $\mathbf{e}^{\mathbf{q},\sigma}=\left[ \mathbf{e}^{\mathbf{q},\sigma}_1,\mathbf{e}^{\mathbf{q},\sigma}_2,...,\mathbf{e}^{\mathbf{q},\sigma}_N \right]$~\cite{kettel,Leibfried},
\begin{equation}
\begin{aligned}
\mathbf{e}_i^{\mathbf{q},\sigma} = \frac{ \mathbf{P}^{\hat{\mathbf{q}},\sigma} }{\sqrt{N}} \exp ( \text{i} \mathbf{q} \cdot \mathbf{r}_i ).
\end{aligned}
\end{equation}
$\mathbf{q}$ is the wave vector, and $\hat{\mathbf{q}} = \mathbf{q}/\left| \mathbf{q} \right|$.
Due to the periodic boundary condition of the finite dimension $L$,
$\mathbf{q}$ is discretized as $\mathbf{q} = ({2\pi}/{L}) (i, j, k )$
for 3D and as $\mathbf{q} = ({2\pi}/{L}) (i, j)$ for 2D ($i,j,k = 0, 1,2,3,...$ are integers).
The value of $\sigma$ denotes one longitudinal ($\sigma=1$) and two
transverse ($\sigma=2,3$) modes for 3D, and it denotes one longitudinal ($\sigma=1$) and one transverse ($\sigma=2$) modes for 2D.
$\mathbf{P}^{\hat{\mathbf{q}},\sigma}$ is a unit vector representing the direction of polarization, which is determined as $\mathbf{P}^{\hat{\mathbf{q}},1} = \hat{\mathbf{q}}$ (longitudinal) and $\mathbf{P}^{\hat{\mathbf{q}},2} \cdot \hat{\mathbf{q}}= \mathbf{P}^{\hat{\mathbf{q}},3} \cdot \hat{\mathbf{q}} = 0$ (transverse).
Note that the vectors $\mathbf{e}^{\mathbf{q},\sigma}$ are orthonormal as $\mathbf{e}^{\mathbf{q},\sigma} \cdot \mathbf{e}^{\mathbf{q}',\sigma'} = \sum_{i=1}^N \mathbf{e}_i^{\mathbf{q},\sigma} \cdot \mathbf{e}_i^{\mathbf{q}',\sigma'} \approx \delta_{\mathbf{q},\mathbf{q}'} \delta_{\sigma,\sigma'}$.
Here, strictly speaking, $\mathbf{e}^{\mathbf{q},\sigma} \cdot \mathbf{e}^{\mathbf{q}',\sigma'} = \delta_{\sigma,\sigma'}$ for $\mathbf{q}=\mathbf{q}'$ and $= \mathcal{O}(N^{-1/2})$ for $\mathbf{q} \ne \mathbf{q}'$, which becomes exactly $\mathbf{e}^{\mathbf{q},\sigma} \cdot \mathbf{e}^{\mathbf{q}',\sigma'} = \delta_{\mathbf{q},\mathbf{q}'} \delta_{\sigma,\sigma'}$ as $N \to \infty$ (the thermodynamic limit).

In the low-$\omega$ limit, the continuum mechanics determine the dispersion relation as $\omega^{\mathbf{q},\sigma} = c^\sigma \left| \mathbf{q} \right|$.
$c^\sigma$ is the phonon speed; $c^1 = c_L = \sqrt{{(K+4G/3)}/{\rho}}$
and $c^2 = c^3 = c_T = \sqrt{{G}/{\rho}}$ for 3D, and $c^1 = c_L =
\sqrt{{(K+G)}/{\rho}}$ and $c^2 = c_T = \sqrt{{G}/{\rho}}$ for 2D.
Here, $\rho$ is the mass density, and $K$ and $G$ are the bulk and shear elastic moduli, respectively.
In this study, we calculate $K$ and $G$ using the harmonic formulation~\cite{Mizuno3_2016}.
Debye theory counts the number of phonons to yield the vDOS as
\begin{equation} \label{equofdebye}
\begin{aligned}
g_D(\omega) = A_0 \omega^{d-1} = \left( \frac{d}{\omega_D^d} \right) \omega^{d-1},
\end{aligned}
\end{equation}
where $A_0 = d/\omega_D^d$ is the Debye level and $\omega_D$ is the
Debye frequency; $\omega_D = \left[ (18 \pi^2
  \hat{\rho})/\left(c_L^{-3} + 2c_T^{-3}\right) \right]^{1/3}$ for 3D,
and $\omega_D = \left[ (8 \pi \hat{\rho})/\left(c_L^{-2} + c_T^{-2}
  \right) \right]^{1/2}$ for 2D, where $\hat{\rho} = N/L^d$ is the number density.
Close to the unjamming transition, $\omega_D \propto c_T \propto
\omega_\ast^{1/2} \propto
p^{1/4}$~\cite{OHern_2003,Silbert_2005,Silbert_2009,Mizuno3_2016}, and
$A_0 \propto \omega_\ast^{-d/2} \propto p^{-d/4}$ (see also the Supplemental Information, Figs.~S6 and~S7).

\subsection{Phonon order parameter}
We evaluate the extent to which the mode $\mathbf{e}^k$ is close to phonons $\mathbf{e}^{\mathbf{q},\sigma}$ by introducing the phonon order parameter $O^k$ as follows.
The eigen vector $\mathbf{e}^k$ can be expanded in a series of phonons $\mathbf{e}^{\mathbf{q},\sigma}$ (Fourier-series expansion) as $\mathbf{e}^k = \sum_{\mathbf{q},\sigma} A^k_{\mathbf{q},\sigma} \mathbf{e}^{\mathbf{q},\sigma}$.
Then, we can calculate the projection onto one particular phonon $\mathbf{e}^{\mathbf{q},\sigma}$ as
\begin{equation}
\begin{aligned}
O^k_{\mathbf{q},\sigma} = \left| A^k_{\mathbf{q},\sigma} \right|^2 \approx \left| \mathbf{e}^{\mathbf{q},\sigma} \cdot \mathbf{e}^k \right|^2.
\end{aligned}
\end{equation}
Here, note that $\sum_{\mathbf{q},\sigma} O^k_{\mathbf{q},\sigma} \approx 1$ since $\mathbf{e}^k \cdot \mathbf{e}^k = 1$.

If the mode $k$ is a phonon, then $\mathbf{e}^k$ is described as the summation of a finite number of large overlapped phonons; $\mathbf{e}^k = \sum_{\mathbf{q},\sigma;\ O^k_{\mathbf{q},\sigma} \ge N_m/(dN-d)} A^k_{\mathbf{q},\sigma} \mathbf{e}^{\mathbf{q},\sigma}$.
Here, we define ``large overlapped" by $O^k_{\mathbf{q},\sigma} \ge N_m/(dN-d)$, i.e., overlapped by the extent of more than $N_m$ modes.
Considering the above, we define the phonon order parameter $O^k$ as
\begin{equation} \label{equofok}
\begin{aligned}
O^k & =  \sum_{\mathbf{q},\sigma;\ O^k_{\mathbf{q},\sigma} \ge N_m/(dN-d)} O^k_{\mathbf{q},\sigma}.
\end{aligned}
\end{equation}
$O^k = 1$ for a phonon, whereas $O^k = 0$ for a mode that is considerably different from phonons.
In the present study, $N_m = 100$ was employed; however, we confirmed
that our results and conclusions do not depend on the choice of the value of $N_m$.

In addition, we calculate the (normalized) spatial correlation function $C^k_{\hat{\mathbf{q}},\sigma}(\left| \hat{\mathbf{q}}\cdot \mathbf{r} \right|)$ of $\mathbf{e}^k$ projected to $\mathbf{P}^{\hat{\mathbf{q}},\sigma}$ as
\begin{equation} \label{equofck}
\begin{aligned}
C^k_{\hat{\mathbf{q}},\sigma}(\left| \hat{\mathbf{q}}\cdot (\mathbf{r}_i - \mathbf{r}_j) \right|) = \frac{\left< \left( \mathbf{P}^{\hat{\mathbf{q}},\sigma} \cdot \mathbf{e}^k_i(\mathbf{r}_i) \right) \left( \mathbf{P}^{\hat{\mathbf{q}},\sigma} \cdot \mathbf{e}^k_j(\mathbf{r}_j) \right) \right>}{\left< \left( \mathbf{P}^{\hat{\mathbf{q}},\sigma} \cdot \mathbf{e}^k_i(\mathbf{r}_i) \right) \left( \mathbf{P}^{\hat{\mathbf{q}},\sigma} \cdot \mathbf{e}^k_i(\mathbf{r}_i) \right) \right>},
\end{aligned}
\end{equation}
where $\left< \right>$ denotes the average over all pairs of particles $i,j$.
If the vibrational mode $\mathbf{e}^k$ is a phonon, then $C^k_{\hat{\mathbf{q}},\sigma}(\left| \hat{\mathbf{q}}\cdot \mathbf{r} \right|)$ exhibits a sinusoidal curve.

\subsection{Participation ratio}
We measure the extent of vibrational localization by the participation ratio $P^k$~\cite{mazzacurati_1996,Taraskin_1999,Schober_2004,Xu_2010},
\begin{equation} \label{equofpk}
P^k = \frac{1}{N} \left[ \sum_{i=1}^N \left( \mathbf{e}_i^k \cdot \mathbf{e}_i^k \right)^2 \right]^{-1}.
\end{equation}
As extreme cases, $P^k = 1$ for an ideal mode where all the particles vibrate equivalently, and $P^k = 1/N$ for a mode involving only one particle.
In the present study, we determine localized modes using the criterion
of $P^k < 10^{-2}$, i.e., modes involving less than $1\%$ of the total particles.

\subsection{Vibrational energy}
Finally, we calculate the vibrational energy $\delta E^{k \parallel}, \delta E^{k \perp}$.
The vector $\mathbf{e}_{ij}^k=\mathbf{e}_{i}^k-\mathbf{e}_{j}^k$ represents the vibrational motion at the contact of particles $i,j$, which can be decomposed to the normal $\mathbf{e}_{ij}^{k \parallel}$ and the tangential $\mathbf{e}_{ij}^{k \perp}$ vibrations with respect to the bond vector $\mathbf{n}_{ij}=(\mathbf{r}_{i}-\mathbf{r}_{j})/\left| \mathbf{r}_{i}-\mathbf{r}_{j} \right|$~\cite{Mizuno3_2016}; $\mathbf{e}_{ij}^{k \parallel} = \left( \mathbf{e}^k_{ij} \cdot {\mathbf{n}_{ij}} \right) \mathbf{n}_{ij}$ and $\mathbf{e}_{ij}^{k \perp} = \mathbf{e}^k_{ij} - \left( \mathbf{e}^k_{ij} \cdot {\mathbf{n}_{ij}} \right) \mathbf{n}_{ij}$.
Accordingly, the vibrational energy $\delta E^k = {\omega^k}^2/2$ can be decomposed as
\begin{equation} \label{equofve}
\begin{aligned}
\delta E^k &= \sum_{(i,j)} \left[ \frac{\phi''(r_{ij})}{2} \left( \mathbf{e}_{ij}^{k \parallel} \right)^2 + \frac{\phi'(r_{ij})}{2 r_{ij}} \left( \mathbf{e}_{ij}^{k \perp} \right)^2 \right], \\
&= \delta E^{k \parallel} - \delta E^{k \perp}.
\end{aligned}
\end{equation}
For the present repulsive system, $\delta E^{k \perp}$ is always positive.
If the mode $k$ is a phonon, then $\delta E^{k \parallel}$ and $\delta E^{k \perp}$ are both proportional to $\delta E^k \propto \omega^2$.
On the other hand, for the floppy mode, the tangential $\delta E^{k \perp}$ exhibits $\omega$-independent behaviour, $\delta E^{k \perp} \propto \omega^0$~\cite{Mizuno3_2016}.

\begin{figure*}[t]
\centering
\includegraphics[width=0.98\textwidth]{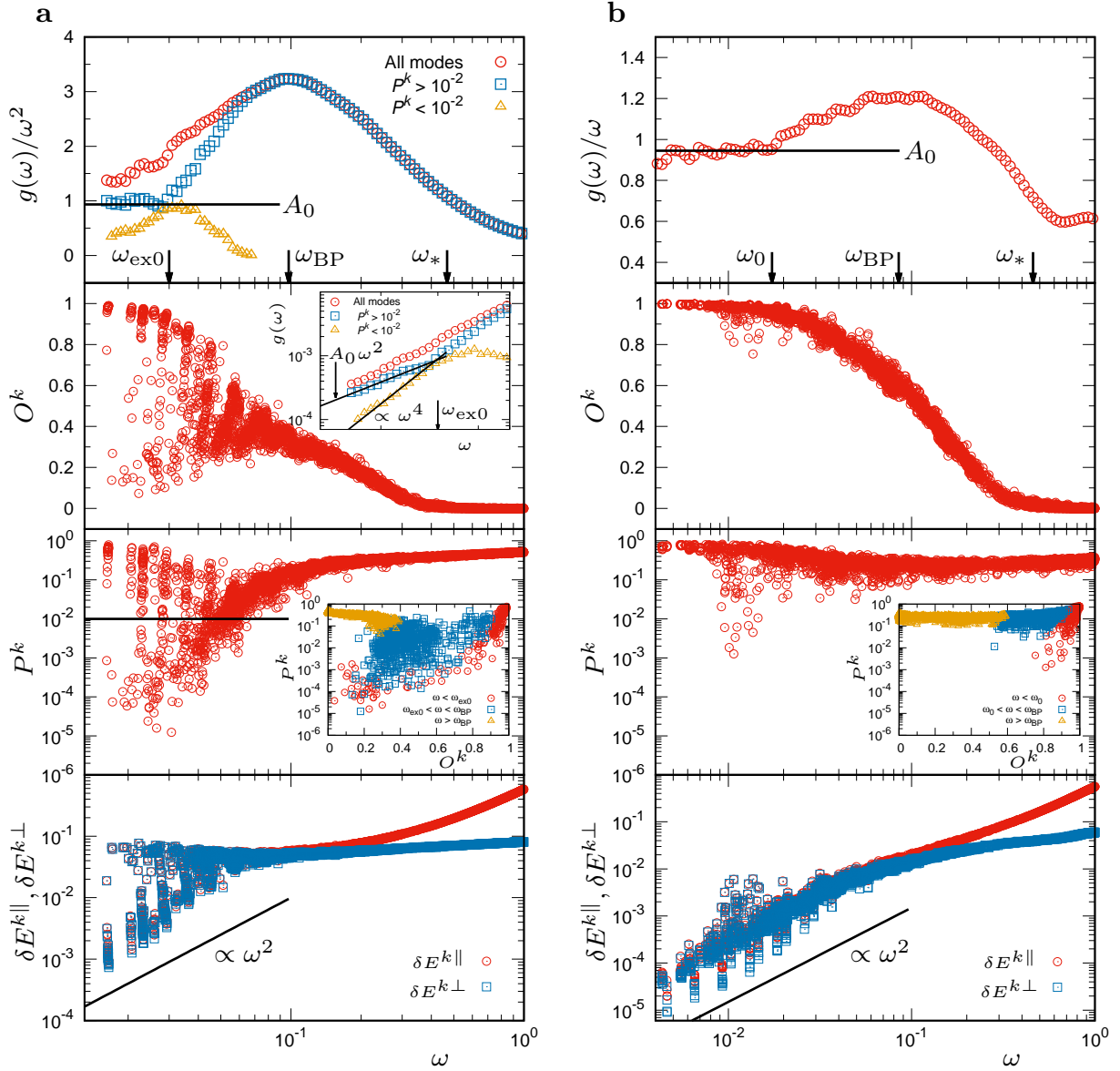}
\vspace*{0mm}
\caption{\label{fig1}
{\bf Vibrational modes in the model amorphous solid.}
Plots of $g(\omega)/\omega^{d-1}$ and of the $O^k$, $P^k$, $\delta
E^{k \parallel}$, and $\delta E^{k \perp}$ of each mode $k$ as functions of $\omega$.
{\bf a} (left panels), 3D model system ($d=3$).
{\bf b} (right panels), 2D model system ($d=2$).
The packing pressure is $p=5 \times 10^{-2}$.
The inset in the third panel from the top provides the plot of $O^k$ versus $P^k$.
For the 3D system in {\bf a}, $g_\text{ex}(\omega)$, the vDOS of
extended modes ($P^k > 10^{-2}$) and $g_\text{loc}(\omega)$, which is
that of localized modes ($P^k < 10^{-2}$), are also presented in the top panel and the inset of the second panel.
}
\end{figure*}

\section{Result}
We first study the vibrational modes of the three-dimensional (3D) model system at a pressure (density) above the unjamming transition (see the Methods section).
Figure~\ref{fig1}a~(top panel, red circles) presents the reduced vDOS $g(\omega)/\omega^2$ at $p=5 \times 10^{-2}$ (packing fraction $\varphi \approx 0.73$) together with the Debye level $A_0$.
The value of $A_0$ is independently determined from the macroscopic mechanical moduli~[Eq.~(\ref{equofdebye})].
The reduced vDOS clearly exhibits a maximum, i.e., the boson peak,
which has not yet been observed for this model~\cite{Silbert_2005,Silbert_2009}.
We place arrows indicating the position of the boson peak
$\omega_\text{BP}$ and of $\omega_\ast$ (the onset frequency of the plateau).
Please refer to the Supplementary Information, Fig.~S1, for the plateau and its onset frequency $\omega^\ast$ in $g(\omega)$.
The value of $\omega_\text{BP}$ is approximately five times smaller than $\omega_\ast$, which is why the boson peak is technically difficult to detect for this model.
Further decreasing the frequency below $\omega_\text{BP}$,
$g(\omega)/\omega^2$ decreases towards but does not reach $A_0$ in the
frequency region that we studied.
We will carefully discuss this point after characterizing the nature of the vibrational modes.

To characterize the modes, we calculate three different parameters~(see the Methods section).
First, the second panel from the top in Fig.~\ref{fig1}a presents the phonon order parameter $O^k$, defined as the projection onto phonons~[Eq.~(\ref{equofok})], for each mode $k$.
$O^k$ measures the extent to which the mode $k$ is close to phonons, taking values from $1$ (phonon) to $0$ (non-phonon).
At $\omega > \omega_\ast$, $O^k$ is nearly zero, which confirms that these modes, called floppy modes (disordered extended modes)~\cite{Wyart_2005,Wyart_2006}, are largely different from phonons.
As the frequency is decreased from $\omega_\ast$ to
$\omega_\text{BP}$, $O^k$ smoothly increases to $\approx 0.3$. 
This result indicates that the modes around the boson peak have a hybrid character of phonons and floppy modes.
Quite remarkably, as the frequency is further decreased below
$\omega_\text{BP}$, the modes are divided into two groups: $O^k$
increases with decreasing frequency in one group, whereas $O^k$
decreases in the other group. 
In the former group, $O^k$ converges to almost $1$ at $\omega_\text{ex0}$ (we provide the precise definition of $\omega_\text{ex0}$ later); namely, these modes are phonons
\footnote{
$O^k$ of these phonon modes are close to but not exactly $1$,
  indicating that they are very weakly perturbed. 
Exact $O^k = 1$ may be realized only in the limit of $\omega \to 0$.}.
Second, the third panel plots the participation ratio $P^k$ that evaluates the extent of spatial localization of the mode $k$~[Eq.~(\ref{equofpk})].
$P^k$ takes values from 1 (extended over all particles equally) to $1/N \ll 1$ (localized in one particle)~\cite{mazzacurati_1996,Taraskin_1999,Schober_2004,Xu_2010}.
As shown in the panel, $P^k$ also exhibits the division of modes into two groups: one approaches $P^k = \mathcal{O}(1)$ with decreasing $\omega$, and the other approaches $P^k = \mathcal{O}(1/N)$.
The inset in this panel shows that the non-phonon modes (small $O^k$) are localized (small $P^k$), whereas the phonon modes (large $O^k$) are extended (large $P^k$) at $\omega < \omega_\text{ex0}$. 
Third, the bottom panel presents the normal and tangential vibrational energies, $\delta E^{k \parallel}, \delta E^{k \perp}$~[Eq.~(\ref{equofve})]~\cite{Mizuno3_2016}.
Again, the modes are split into two groups. 
The phonon modes follow the scaling behaviour $\delta E^{k \parallel} \approx \delta E^{k \perp} \propto \omega^2$ as in the case of crystalline solids, whereas the non-phonon localized modes are characterized by the $\omega$-independent behaviour of $\delta E^{k \perp}$ as in the case of the floppy modes at $\omega > \omega_\ast$.
These three coherent results demonstrate that the phonon modes and the non-phonon localized modes coexist at $\omega < \omega_\text{ex0}$.

\begin{figure*}[t]
\centering
\includegraphics[width=0.99\textwidth]{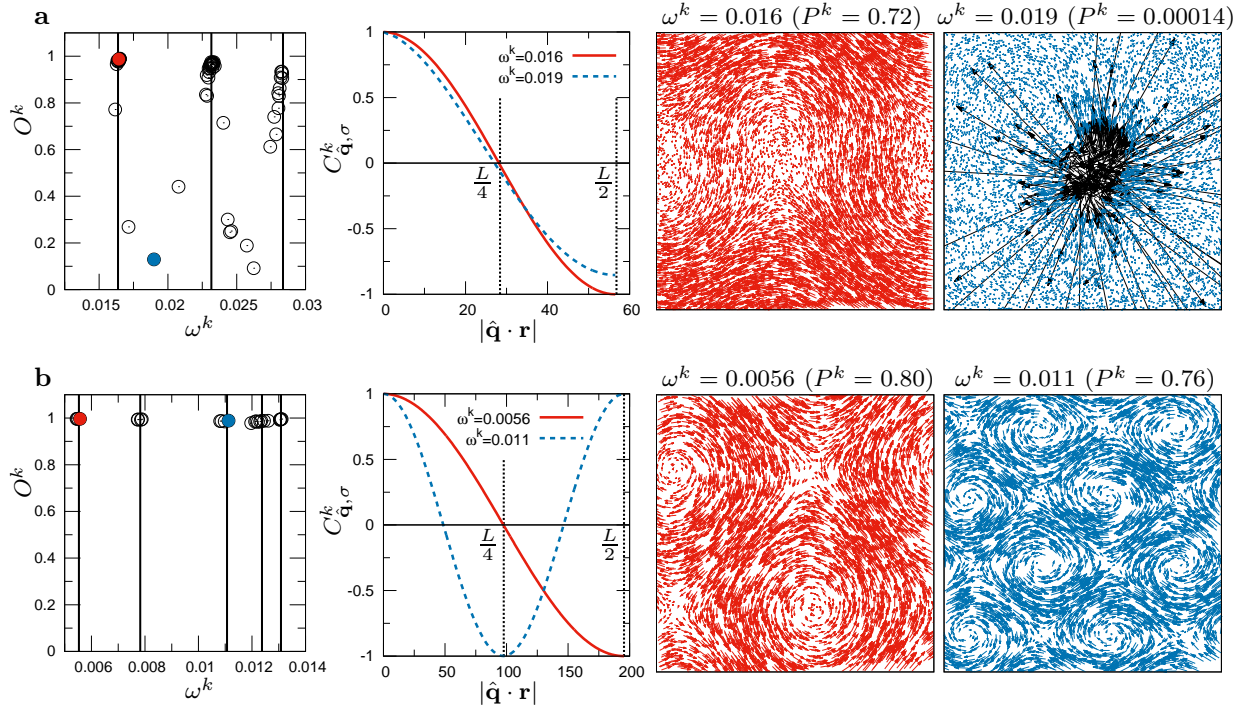}
\vspace*{0mm}
\caption{\label{fig2}
{\bf Vibrational modes in the low-$\boldsymbol{\omega}$ regime.}
{\bf a} (upper panels), ${\omega < \omega_{\text{ex0}}}$ of the 3D model system.
{\bf b} (lower panels), ${\omega < \omega_0}$ of the 2D model system.
$p=5 \times 10^{-2}$.
The system size is $N=2048000$, $L=114$ (3D) and $N=128000$, $L=391$ (2D).
The leftmost panel plots $O^k$ versus $\omega^k$ for each mode $k$ (symbols) and the energy levels of phonons (vertical lines). 
The remaining panels provide more detailed information on the two modes highlighted by red and blue closed circles in the left panel.
The second panel from the left presents the spatial correlation function of the eigen-vector field $\mathbf{e}^k$, $C^k_{\hat{\mathbf{q}},\sigma} (\hat{\mathbf{q}} \cdot \mathbf{r})$ along the $[100]$ transverse wave. 
The third and fourth panels visualize the eigen-vector fields of these two modes.
For the 3D system in {\bf a}, the vector fields are plotted at a fixed plane of a thickness of particle diameter, and the localized region is emphasized by black arrows.
}
\end{figure*}

We now perform more stringent tests of the nature of these two types of modes. 
In Fig.~\ref{fig2}a (left panel), the phonon order parameter $O^k$ is plotted against the frequency $\omega^k$ for each mode $k$ in the system of $N=2048000$.
Because the system is in a finite box, phonons should have discrete energy levels. 
We calculate these energy levels from the macroscopic elastic moduli
and the linear dimension of the box $L=114$~(see the Methods section),
and we display them as vertical lines in the figure.
Indeed, the phonon modes (modes with large $O^k$) sit on these levels. 
Conversely, the soft localized modes (modes with small $O^k$) are in the gaps between the different levels.
Furthermore, the right panels show the eigen-vector field $\mathbf
e^k$ of the representative modes (highlighted as filled circles in the
left panel), which demonstrates that the mode on the level has a
phonon structure, whereas the other mode is localized. 
These results unambiguously establish the distinction between phonon modes and soft localized modes. 

However, note that the soft localized modes are not truly localized.
This is best observed in the spatial correlation
function~$C^k_{\hat{\mathbf{q}},\sigma}(\left| \hat{\mathbf{q}}\cdot
\mathbf{r} \right|)$~[Eq.~(\ref{equofck})], as shown in the middle panel of Fig.~\ref{fig2}a.
Here, we calculate the spatial correlation of the eigen-vector field $\mathbf{e}^k$ along the $[100]$ transverse wave~(see the Methods section). 
This function exhibits a nice sinusoidal shape not only for the phonon mode but also for the localized mode.
This result indicates that the localized mode has disordered
vibrational motions in the localized region; however, these motions are accompanied by extended phonon vibrations in the background.
Consistent with this result, we observe that the participation ratios of these modes do not scale as $1/N$ with increasing system size $N$ at a fixed $\omega$.
Therefore, these modes are quasi-localized~\cite{mazzacurati_1996,Schober_2004}.
This feature is very similar to the modes in defect crystals, where the quasi-localized modes are produced by hybridization of the extended phonon and the localized defect modes~\cite{Leibfried,Schober_2004}.

A clear distinction between phonon modes and soft localized modes enables us to consider the vDOSs of these two types of modes separately.
We define $g_\text{ex}(\omega)$ as the vDOS of modes with $P^k >
10^{-2}$ and $g_\text{loc}(\omega)$ as the vDOS of those with $P^k < 10^{-2}$, and
we plot them in Fig.~\ref{fig1}a (the top panel and the inset that is
the second from the top).
As shown, $g_\text{ex}(\omega)$ converges exactly to the Debye behaviour $A_0 \omega^2$ at a finite $\omega$, which we define as $\omega_{\text{ex0}}$.
However, $g_\text{loc}(\omega)$ follows a different scaling law $g_\text{loc}(\omega) \propto \omega^4$. 
Thus, we now conclude that the phonon modes following the Debye law $g_\text{ex}(\omega) = A_0 \omega^2$ and the soft localized modes following the other law $g_\text{loc}(\omega) \propto \omega^4$ coexist at $\omega < \omega_\text{ex0}$.
This result then suggests that the full vDOS $g(\omega) = g_\text{ex}(\omega) + g_\text{loc}(\omega)$ eventually converges to the Debye vDOS in the limit of $\omega \to 0$ because $g_\text{loc}(\omega) \propto \omega^4$ decays faster than $g_\text{ex}(\omega) \propto \omega^2$. 
Here we note that the $\omega^4$ scaling is the same law proposed in the soft-potential model~\cite{Karpov_1983,Buchenau_1991,Buchenau_1992,Gurevich_2003}. 
Also, it was recently reported that the $\omega^4$ law can be observed in the Heisenberg spin-glass and structural glasses if the effects of phonons are suppressed by introducing a random potential~\cite{Baity-Jesi2015}, tuning the system size to be sufficiently small~\cite{Lerner_2016}, or focusing on the low-frequency regime below the lowest phonon mode~\cite{Gartner_2016}.
In the present work, for the first time to our knowledge, we found that the vibrational modes are spontaneously divided into the phonon modes and the soft localized modes at $\omega < \omega_{\text{ex0}}$ and observed that these soft localized modes exactly follow the $\omega^4$ law.

\begin{figure}[t]
\centering
\includegraphics[width=0.49\textwidth]{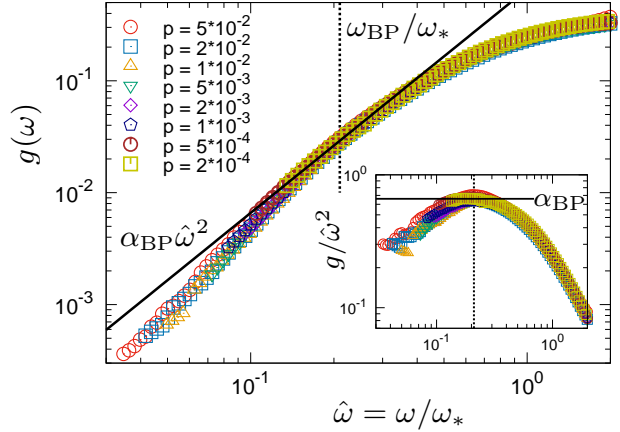}
\vspace*{0mm}
\caption{\label{fig3}
{\bf The vDOSs at different $\boldsymbol{p}$ in the 3D model system.}
$g(\omega)$ is plotted against the scaled frequency $\hat{\omega}=\omega/\omega_\ast$ for several different $p$.
The inset is the same as the main panel but for the reduced vDOSs $g/\hat{\omega}^2$.
Around $\omega_\text{BP} = 0.21 \omega_\ast$, the vDOSs are collapsed onto the non-Debye scaling law $g(\omega) = \alpha_\text{BP} \hat{\omega}^2$ with $\alpha_\text{BP} = 0.66$.
}
\end{figure}

We repeated this analysis with different packing pressures (densities). 
We observe that the basic features are unchanged (see the Supplementary Information, Fig.~S2a). 
Furthermore, we find that the vDOSs at different pressures can be summarized by the scaling laws. 
To illustrate this result, we determined $\omega_\ast$ for each pressure and introduced the scaled frequency $\hat{\omega} = \omega/\omega_\ast$. 
Here, we confirmed the well-established property of $\omega_\ast \propto p^{1/2}$~\cite{Silbert_2005,Silbert_2009,Wyart_2005,Wyart_2006}.
Then, we plot $g(\omega)$ and $g(\omega)/\hat{\omega}^2$ against $\hat{\omega}$ at various pressures in Fig.~\ref{fig3}. 
All the data perfectly collapse around and above the boson peak.
We emphasize that there are no adjustable parameters for this collapse. 
At $\hat{\omega} \gtrsim 1$, i.e., $\omega \gtrsim \omega_\ast$, $g(\omega)$ exhibits a constant plateau $g(\omega) = \alpha_\ast = 0.37$~ (see the Supplementary Information, Fig.~S1a). 
The boson peak is located at $\hat{\omega} = 0.21$; thus, $\omega_\text{BP} = 0.21 \omega_\ast$ at all pressures. 
The vDOS around the boson peak can be fitted to the non-Debye scaling $\alpha_\text{BP} \hat{\omega}^2$ with $\alpha_\text{BP} = 0.66$, as predicted by the mean-field theory~\cite{DeGiuli_2014,Franz_2015} and observed in simulations~\cite{Charbonneau_2016}.
However, the data systematically deviate from this scaling law around $\hat{\omega} \approx 0.1$ at all pressures. 
This result can be contrasted with the results of replica theory, which predicts that the present system has a marginally stable glass phase in some finite region of the density above the unjamming point~\cite{Biroli_2016}, and the region of the non-Debye scaling law $\alpha_\text{BP} \hat{\omega}^2$ extends down to $\omega \to 0$ in this phase~\cite{Franz_2015}.
Our result is more consistent with the results in Refs.~\cite{DeGiuli_2014,Lerner_2014}. 

\begin{figure}[t]
\centering
\includegraphics[width=0.49\textwidth]{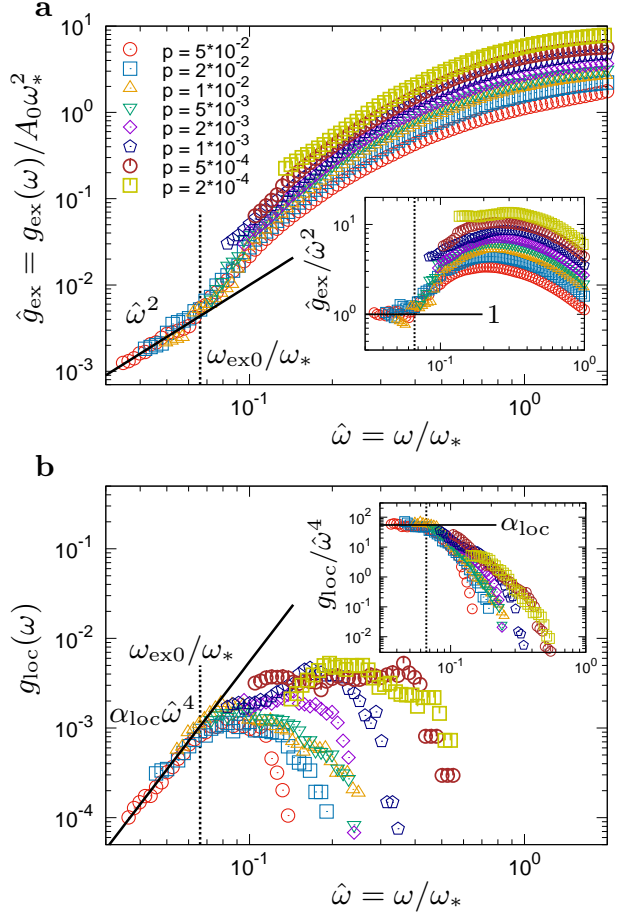}
\vspace*{0mm}
\caption{\label{fig4}
{\bf The vDOSs of the extended modes and the soft localized modes in
  the 3D model system.}
{\bf a}, Plot of the scaled vDOSs of the extended modes ($P^k>10^{-2}$), $\hat{g}_\text{ex} = g_\text{ex}(\omega)/A_0 \omega_\ast^2$, against the scaled frequency $\hat{\omega}=\omega/\omega_\ast$.
{\bf b}, Plot of the vDOSs of the localized modes ($P^k<10^{-2}$), $g_\text{loc}(\omega)$, against $\hat{\omega}$.
The insets are the same as the main panels but for the reduced vDOSs $\hat{g}_\text{ex}/\hat{\omega}^2$ in {\bf a} and ${g}_\text{loc}/\hat{\omega}^4$ in {\bf b}.
Below $\omega_\text{ex0}=0.066 \omega_\ast$, the vDOSs collapse onto $g_\text{ex}(\omega) = A_0 \omega^2$ and $g_\text{loc}(\omega) = \alpha_\text{loc} (\omega/\omega_\ast)^4$ with $\alpha_\text{loc} = 58$.
}
\end{figure}

Instead, at lower frequency $\omega \ll \omega_{BP}$, we find that another scaling law works. 
In Fig.~\ref{fig4}a, we plot $g_\text{ex}(\omega)/A_0 \omega_\ast^2$ against $\hat{\omega}$.
We note that the Debye level $A_0$ is related to $\omega_\ast$ as $A_0 = \alpha_0 \omega_\ast^{-3/2}$ with $\alpha_0 = 0.27$. 
When decreasing $\hat{\omega}$, all the data at different pressures collapse and converge to $g_\text{ex}(\omega)/A_0 \omega_\ast^2 = \hat{\omega}^2$, which is exactly the Debye behaviour $g_\text{ex}(\omega) = A_0 \omega^2$.
This convergence occurs at $\hat{\omega} = 0.066$; thus, $\omega_{\text{ex0}} = 0.066 \omega_\ast$ at all pressures. 
Next, in Fig.~\ref{fig4}b, we plot $g_\text{loc}(\omega)$ against $\hat{\omega}$.
Remarkably, all the data converge to another universal scaling law $g_\text{loc}(\omega) = \alpha_\text{loc} \hat{\omega}^4$ with $\alpha_\text{loc} = 58$.
This convergence occurs at $\hat{\omega} = 0.066$ as in $g_\text{ex}(\omega)$.
These two results demonstrate that the full vDOS $g(\omega) = g_\text{ex}(\omega) + g_\text{loc}(\omega)$ can be expressed as $g(\omega) = A_0 \omega^2 + \alpha_\text{loc}\hat{\omega}^4$ at $\omega \lesssim \omega_{\text{ex0}}$. 

Therefore, by collecting the results in all the frequency regions, we
can write the functional form of the vDOS that covers the continuum
limit as follows:
\begin{equation} \label{dos}
g(\omega) = 
\left\{ \begin{aligned}
& \alpha_\ast & (\omega \gtrsim \omega_\ast ), \\
& \alpha_\text{BP} \left( \frac{\omega}{\omega_\ast} \right)^2 & (\omega \sim \omega_\text{BP}), \\
& A_0 \omega^2 + \alpha_\text{loc} \left( \frac{\omega}{\omega_\ast} \right)^4 & (\omega \lesssim \omega_\text{ex0}),
\end{aligned} \right. 
\end{equation}
with $\omega_\text{BP} = 0.21 \omega_\ast$, $\omega_\text{ex0} = 0.066 \omega_\ast$, $\alpha_\ast = 0.37$, $\alpha_\text{BP} = 0.66$, $\alpha_\text{loc} = 58$, and $A_0 = \alpha_0 \omega_\ast^{-3/2}$ with $\alpha_0 = 0.27$.
Strikingly, except for the phonon part $A_0 \omega^2 = \alpha_0 \omega_\ast^{1/2}(\omega/\omega_\ast)^2$, the vDOS takes the form of a universal function of the reduced frequency $\omega/\omega_\ast$ only.
In other words, the non-phonon contribution to the vDOS can be expressed as $g_\text{non-phonon}(\omega) = G({\omega}/{\omega_\ast})$, where $G(x) = \alpha_\ast$ for $x \gtrsim 1$, $G(x) = \alpha_\text{BP} x^2$ for $x \sim 0.21$, and $G(x) = \alpha_\text{loc} x^4$ for $x \lesssim 0.066$. 
This result implies that all the non-phonon vibrations, including the soft localized modes, are controlled by the physics of $\omega_\ast$, namely, the isostaticity and the marginal stability. 
We note that the soft potential model predicts $g_\text{loc}(\omega) \approx \omega_\text{BP}^{-3} \omega^4$~\cite{Gurevich_2003}; the exponent of $\omega$ is consistent with our result, but the coefficient is not. 

To further discuss the origin of the non-phonon behaviours, we perform
a vibrational mode analysis of the ``unstressed'' system. 
The unstressed system is defined as the system in which the particle-particle contacts of the original system are replaced with relaxed springs~(see the Methods section). 
In the present model, the unstressed system is known to be far from the marginally stable state~\cite{Wyart_2006,DeGiuli_2014,Lerner_2014}.
Thus, by observing whether a mode disappears in the unstressed system,
one can evaluate whether the mode originated from the marginal stability.
We observe that the scaling region for $g(\omega) = \alpha_\text{BP}
\hat{\omega}^2$ is suppressed in the unstressed system, which confirms
that these modes originated from the marginal stability~\cite{DeGiuli_2014,Franz_2015}. 
Furthermore, the soft localized modes are strongly quelled~(see the Supplementary Information, Figs.~S2b and~S4)
\footnote{
We observe that the unstressed system begins to show soft localized modes when the system is very close to the unjamming transition. 
This may be because even the unstressed system is brought near the elastic instability when approaching the unjamming transition.
}.
This result suggests that the soft localized modes also originated from the marginal stability, although they are not captured by the current mean-field framework.

\begin{figure}[t]
\centering
\includegraphics[width=0.49\textwidth]{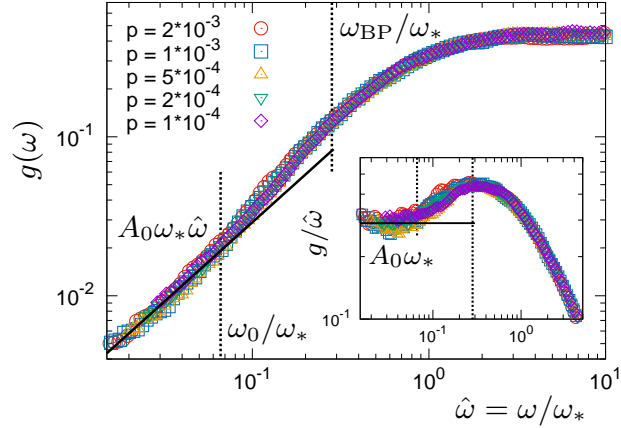}
\vspace*{0mm}
\caption{\label{fig5}
{\bf The vDOSs at different $\boldsymbol{p}$ in the 2D model system.}
$g(\omega)$ is plotted against the scaled frequency $\hat{\omega}=\omega / \omega_\ast$ for several different $p$.
The inset is the same as the main panel but for the reduced vDOSs $g/\hat{\omega}$.
The full vDOSs are collapsed onto a universal function of the scaled
frequency $\hat{\omega}$ over the entire $\omega$ regime.
At $\omega_\text{0}=0.066\omega_\ast$, $g(\omega)$ converges to the Debye vDOS $g(\omega) = A_0 \omega$.
}
\end{figure}

Finally, we focus on the vibrational modes of the two-dimensional (2D) model system, where we encounter a surprisingly different situation.
Figure~\ref{fig1}b shows that $g(\omega)$ converges smoothly and exactly to the Debye vDOS at a finite frequency that we define as $\omega_0$.
Below $\omega_0$, the modes are characterized by $O^k \approx 1$; namely, these modes are phonons.
Therefore, no group of soft localized modes appears in 2D. 
This result is more evident in Fig.~\ref{fig2}b; all the vibrational modes at $\omega < \omega_0$ sit on the energy levels of phonons, and they also have the spatial structures of phonons.
Another interesting feature is that the full vDOSs at various pressures are expressed as a universal function of $\omega/\omega_\ast$ over the entire $\omega$ regime, as demonstrated in Fig.~\ref{fig5}. 
This result can be rationalized by observing that the Debye (phonon) vDOS is $g(\omega) = A_0 \omega^{d-1}$ with $A_0 \propto \omega_\ast^{-d/2}$; thus, it also becomes a universal function of $\omega/\omega_\ast$ in 2D, as does the non-phonon contribution $g_\text{non-phonon}(\omega)$.
Similar convergence to phonons and collapses of the vDOSs were reported for 2D Lennard-Jones systems by previous works~\cite{tanguy_2002,shintani_2008}. 
The collapse also indicates that the boson peak amplitude scaled by the Debye level, $g(\omega_\text{BP})/A_0 \omega_\text{BP}^{d-1}$~\cite{shintani_2008}, does not depend on the packing pressure $p$ in 2D. 
In contrast, in 3D, this quantity diverges as $\propto \omega_\ast^{-1/2}$ at the unjamming transition, as shown in Eq.~(\ref{dos}) and in Fig.~\ref{fig4}a (see also the Supplementary Information, Eq.~(S3), Figs.~S6 and~S7). 
We also note that the recent work~\cite{Lerner_2016} indicated the $\omega^4$ law of non-phonon modes can be observed even in 2D amorphous systems if the effects of phonons are suppressed.

\section{Conclusion}
In conclusion, we have used a large-scale numerical simulation to
observe the continuum limit of the vibrational modes in a model amorphous solid. 
In 3D, we have found that the vDOS follows the non-Debye scaling
$g(\omega) = \alpha_\text{BP}(\omega/\omega_\ast)^2$ only around and
above the boson peak, and below the boson peak, the vibrational modes
are divided into two groups: the modes in one group converge to the
phonon modes following the Debye law $g_\text{ex}(\omega) = A_0 \omega^2$,
and those in the other group converge to the soft localized modes following another universal non-Debye scaling $g_\text{loc}(\omega) = \alpha_\text{loc} (\omega/\omega_\ast)^4$.
Strikingly, all the non-phonon contributions to the vDOSs at different pressures can be expressed as a universal function of the reduced frequency $\omega/\omega_\ast$. 
In contrast, completely different behaviours are observed in 2D:
vibrational modes smoothly converge to phonons without the appearance of the group of soft localized modes.

Our results, on the one hand, provide a direct verification of the basic assumption of the soft potential model. 
We showed the coexistence of phonon modes and soft localized modes,
which is the central idea for explaining the low-$T$ anomalies of thermal conduction in this phenomenological model~\cite{Anderson_1972,Karpov_1983,Buchenau_1991,Buchenau_1992,Gurevich_2003}.
On the other hand, the violation of the non-Debye scaling at the boson
peak, the emergence of soft localized modes with another non-Debye
scaling, and the crucial difference between 2D and 3D appear to be beyond the reach of the current mean-field theory.
However, the fact that the non-phonon contributions to the vDOS are expressed as a universal function of $\omega/\omega_\ast$ suggests that these features are linked to the isostaticity and the marginal stability, which are captured by the mean-field theory~\cite{Wyart_2005,Wyart_2006,Parisi2010,Wyart_2010,Berthier2011a,Wyart2012,DeGiuli_2014,Charbonneau2014b,Charbonneau2014,Franz_2015,Biroli_2016}.

\section*{Acknowledgments}
We thank H.~Ikeda, Y.~Jin, L.E.~Silbert, P.~Charbonneau, F.~Zamponi, L.~Berthier, E.~Lerner, E. Bouchbinder, and K.~Miyazaki for useful discussions and suggestions.
The numerical calculations were partly carried out on SGI Altix ICE XA at Institute for Solid State Physics, The University of Tokyo, Japan.

\section*{Author contributions}
H.M. and A.I. conceived the project. 
H.M., H.S., and A.I. carried out numerical simulations. 
H.M. and A.I. wrote the paper.

\section*{Additional information}
Supplementary information is available.
Correspondence and requests for materials should be addressed to H.M.~(hideyuki.mizuno@phys.c.u-tokyo.ac.jp) or A.I.~(atsushi.ikeda@phys.c.u-tokyo.ac.jp).

%

\bibliographystyle{apsrev4-1}
\bibliography{paper}

\end{document}